\documentclass[10pt, onecolumn]{article}
\usepackage{multirow}
\usepackage{subfigure}
\usepackage{graphicx}
\usepackage{longtable}
\usepackage{courier}
\usepackage{float}
\usepackage{cite}
\usepackage{amsmath,amssymb}
\usepackage{bbold}
\usepackage{dsfont}
\usepackage[in]{fullpage}
\usepackage[english]{babel}
\usepackage{amsthm}
\usepackage{changes}

\newtheorem{theorem}{Theorem}
\newtheorem{corollary}{Corollary}[theorem]
\newtheorem{definition}{Definition}
\newtheorem*{remark}{Remark}
\newtheorem{lemma}[theorem]{Lemma}
\usepackage{wrapfig}
\usepackage[T1]{fontenc}
\usepackage{authblk}
\usepackage{array}
\usepackage{url}

\usepackage[pdfborder={0 0 0}]{hyperref}
\usepackage[margin=0.55in]{geometry}
\newcommand{\ou}{%
  \mathrel{%
    \vcenter{\offinterlineskip
      \ialign{##\cr$<$\cr\noalign{\kern-1.5pt}$>$\cr}
			 }%
  }%
}	
 \vspace{-8ex}
  \date{}

\begin{document}

 \title{Information-Theoretic Approaches to Differential Privacy}
\author{%
Ay\c{s}e \"{U}nsal \qquad Melek \"{O}nen\\ EURECOM, France \\ \url{firstname.lastname@eurecom.fr}
 %\and 
%Melek \"{O}nen\\ Eurecom, France\\ \url{melek.onen@eurecom.fr}
}

  %\author*[1]{Ay\c{s}e \"{U}nsal}
%%
  %\author[2]{Melek \"{O}nen}
%%
  %\affil[1,2]{EURECOM, France E-mail: firstname.lastname@eurecom.fr}}
%%
  %\affil[2]{EURECOM, France E-mail: melek.onen@eurecom.fr}
%
  %\title{\huge A Statistical Threshold to Remain Undetected for Laplacian and Gaussian Differentially Private Mechanisms}
%
  %\runningtitle{A Statistical Threshold to Remain Undetected for Laplacian and Gaussian Differentially Private Mechanisms}
%
  %%\subtitle{...}
\maketitle
 \pagestyle{plain}

  \begin{abstract}
This tutorial studies relationships between differential privacy and various information-theoretic measures by using several selective articles. 
In particular, we present how these connections can provide new interpretations for the privacy guarantee in systems that deploy differential privacy in an information-theoretic framework.
To this end, the tutorial provides an extensive summary on the existing literature that makes use of information-theoretic measures and tools such as mutual information, min-entropy, Kullback-Leibler divergence and rate-distortion function for quantification and characterization of differential privacy in various settings.
  %\begin{keywords}
	%differential privacy, hypothesis testing, Laplace dp noise, Gaussian dp noise
	%\end{keywords}
	
	\end{abstract}
\section{Introduction \label{sec:intro}}
Over the past decade, machine learning (ML) algorithms have found application in a vast and rapidly growing number of systems for analysing and classifying large amounts of data. Despite the improvement and comfort that was brought to our daily lives by applications that employ these algorithms, they also gave cause for concern in terms of security and data privacy due to their undesired consequences. The increasing popularity of ML techniques opened the door for attackers, especially when these techniques were deployed to be used in critical areas as intrusion detection, autonomous driving or healthcare. In particular, an adversary may look for means to modify the model, misclassify some inputs and consequently succeed in unauthorized cyber-access, car accidents or even health problems. It is not unrealistic to imagine the scenario, where a self-driving car causes an accident due to ignoring a stop sign, which through tampering by an adversary was made to look like a parking sign.

In addition to the security aspect of such an attack, user-data privacy is also prone to violations in this problem. Such data is considered as highly sensitive, since it contains information on location that could lead to discovery of personal habits and may enable vehicle identification. In general, the high quality and high accuracy of ML predictions strongly depend on the collection of large datasets. Such a large-scale data collection gives cause for privacy concerns and makes users vulnerable to fraudulent use of personal information. When individuals willingly share some of their personal data with an Internet service, statistical independence of the representation of the data and the actual individual is a desired quality of the underlying system. At least from a conceptual perspective, a measure of this independence relates to the amount of privacy an individual can expect from the system. However, it is possible to successfully de-anonymize or re-identify the owner of the data as proven by a number of studies as follows \cite{S02, NS08, EJA+11, ZB11}. For instance, Facebook and Cambridge Analytica are real-life examples of massively used online services, which were proven to be a threat to privacy of individuals back in 2010, when Cambridge Analytica acquired a great number of Facebook users' data for the purpose of using the right political advertisement. More recently, it was discovered that Pegasus spyware has been used for reading text messages, tracking calls and locations, accessing the targeted device's camera and microphone in many versions of Apple's iOS and Android \cite{C21}. These few examples of privacy rights' violations make it clear that protecting privacy of personal data is a major concern in today's world. 

In order to address data privacy requirements in such contexts, two application methods are used in current systems, namely local and global privacy. In local privacy methods, individuals publish private version of their own information, as is the case of a social networking website. Global privacy methods make use of a trusted (central) server or curator which publishes private query responses related to a group of individuals. A common characteristic of both approaches is that data is typically coded using some randomizing function prior to its publication. Differential privacy \cite{D06} is a stochastic measure of privacy which is now used in conjunction with ML algorithms while managing large datasets to ensure data privacy of individual users. It has furthermore been used to develop practical methods for protecting private user-data when they provide information to the ML system. In these cases, the use of a differential privacy measure aims to preserve the accuracy of the ML model without incurring a cost of the privacy of individual participants.  %This definition is also applicable to aggregate information of all participants. 
An embedded application in Google's Chrome Web Browser \cite{EPK14}, a Census Bureau project called OntheMAP \cite{MKA+08}, LinkedIn and Apple's iOS 11 are only a few examples of real-life applications which have already deployed differential privacy to address and overcome this vulnerability of users in terms of privacy of personal information. 

\begin{wrapfigure}{r}{0.48\textwidth}		
  \begin{center}	
    \includegraphics[width=0.48\textwidth]{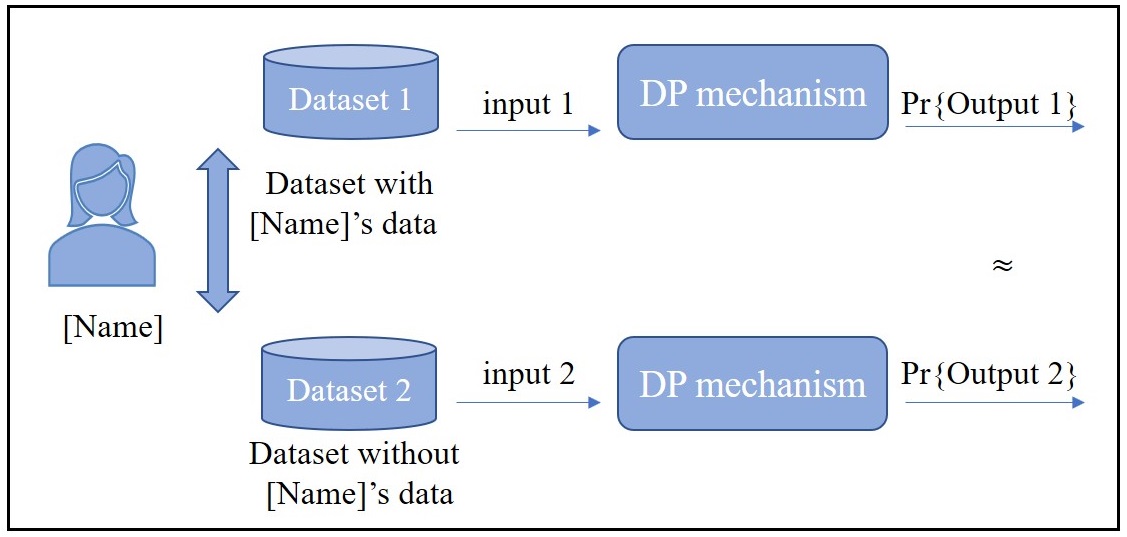}
  \end{center}
  \caption{Differential privacy	\label{fig:dp_def2}}
\end{wrapfigure}
A mechanism or a randomized function of a dataset is called \textit{differentially private} if the absence or presence of any participant's data has a negligible impact on the output of the mechanism when any of the participants decides to submit or equivalently remove their data from a statistical dataset. This idea is roughly depicted in Figure \ref{fig:dp_def2}. 
In some sense, differential privacy is a notion of robustness against such changes in the dataset. The degree of this change is measured and determined by an adjustable privacy parameter (or the privacy budget) and the amount of the change that any single argument to the system reflects on its output is called the sensitivity of the system. The major challenge is to offset the accuracy of the output of a statistical dataset against the level of the privacy protection guaranteed to the participants. Indeed, noisier data results in a stronger level of privacy due to increased randomness and this reflects as a reduction in accuracy of the output. 

\paragraph{Timeliness and necessity of the tutorial:} Differential privacy arises a great interest among researchers particularly from computer science and statistics circles who contributed to what we already know about this strong mathematical formulation of privacy. There are several detailed surveys of what is known today regarding differential privacy from the perspective of computer scientists and statisticians \cite{D08, DS09, DR05}. More recently, also researchers in information theory/electrical engineering circles contributed to the literature on the subject. However, a full information-theoretic understanding of differential privacy and its information-theoretic connections with other trustworthy features are still lacking. This tutorial provides a \textit{selective} summary of what we know regarding the relationship between differential privacy and information theory to enable information theorists, primarily, to build up on that to produce fundamental formulations and limits of privacy in various settings.

\paragraph{On the relevance of information-theoretic connections with differential privacy:} Originally, the use of mutual information functional as privacy metric dates back to \cite{LHA2} for studying the domain of genome privacy prior to the existence of differential privacy. Even though there are different opinions on the form of the exact relation, a number of studies relate the (conditional \cite{CY16} or unconditional \cite{WYZ16, M12}) mutual information between the entries of the dataset and the query response to differential privacy, which could be interpreted as a measure of utility as well as of privacy. Under certain conditions, differential privacy and the \textit{mutual information differential privacy}, have proven to be equal in \cite{CY16} where the authors redefine well-known information theoretic quantities as privacy constraint. Overall, a mutual information-based approach to differential privacy will allow many rules and properties that apply to the mutual information functional to be carried on to differential privacy leaving no room for ambiguity regarding the essence of the privacy guarantee. Furthermore, in \cite{CY16}, the mutual information-based differential privacy removes the requirement for neighborhood among datasets and strengthens the original definition.
Hereafter, we enlist possible directions of research where the information-theoretic connections with differential privacy is pertinent. The reader should note that the following list is exemplary and non-exhaustive. Some items will be studied in detail within the content of this tutorial in further sections.
\begin{itemize}
%%%%%%%%%%OLD VERSION
\item \textbf{Cryptography:} A major example is the connection with \textit{semantic security} via an information-theoretic approach. \cite{BTV12} proves an equivalence between a mutual information based differential privacy constraint and semantic security where a maximization is taken over database distributions. Additionally, \cite{WXH21} introduces a new data-privacy protection model that aims to achieve \textit{Dalenius' goal} as well as to have better utility. The privacy channel capacity results are obtained through direct translations of well-known information theoretic approaches to differential privacy. In particular, the parallel drawn between the information privacy model and the multiple-access channel makes a great promise for the use of an information-theoretic framework to quantify the privacy guarantee that a differentially private system can provide to its users. 
\item \textbf{Security:} \cite{r17} presented an application of the so-called Kullback-Leibler differential privacy \cite{CY16} (to be defined later) for detecting misclassification attacks in differentially private Laplace mechanisms. Accordingly, the corresponding distributions of relative entropy are considered as the differentially private noise with and without the adversary's advantage in order to establish the relationship between the impact of the attack and the detection of the adversary as a function of the sensitivity and the privacy budget of the mechanism. Besides adversarial classification, information-theoretic approaches for bounding the \textit{communication complexity} of computing a function, which originally uses combinatorial measures \cite{Pankratov-thesis} can also be applied to differential privacy. Information complexity \cite{BJK+04} is a lower bound on communication complexity that is obtained using Shannon's mutual information and refers to the minimum amount of information that a communication protocol leaks about its users' inputs. \cite{MMP+10} introduces an upper bound on the information cost of a two-party differentially private protocol using the same approach that will be studied in detail in Section \ref{sec:bounds}. \cite{LKS19}, on the other hand covers, the privacy of physical layer for a two receiver broadcast channel through analyzing connections between a differential privacy based metric to physical layer secrecy. Accordingly, the authors show that for the privacy of anonymous communication networks in the case of a degraded two-user broadcast channel, differentially private receiver-message unlinkability is equivalent up to a constant to several secrecy metrics. Finally, \cite{LKS19} presents the rate region of the $(\epsilon, \delta)-$ differentially private receiver-message unlinkability satisfying strong secrecy. 
\item \textbf{ML:} Probably approximately correct (PAC) learning theory, which composes the mathematical framework of ML, is related to differentially private learning by using mutual information function in \cite{M12+}. Accordingly, the author establishes an information-theoretic connection between Gibbs estimator which gives the minimum of PAC-Bayesian bounds and the exponential mechanisms to show that Gibbs estimator minimizes the expected empirical risk and the mutual information between the sample and the predictor.
\item \textbf{Quantum computation:} There also has been a serious effort towards building connections between quantum computation and differential privacy \cite{HR22, AS19, ZM17, MM17}. Some works build the bridge between the two via \textit{quantum information theory} that draws Shannon information theory, quantum mechanics and computer science together. Quantum differential privacy is originally defined in \cite{ZM17} for adaptation of differential privacy to quantum information processing. \cite{HR22} focuses on quantum differential privacy using an information-theoretic framework, which is translated into quantum divergence. 
\end{itemize} 

\paragraph{Outline:} Section \ref{sec:preliminaries} provides necessary preliminaries from the literature on differential privacy. Introductory preliminaries are followed by novel metrics derived through information-theoretic measures for quantifying privacy guarantee of differentially private mechanisms in Section \ref{sec:shannon_inf} along with their ordering and comparisons. Section \ref{sec:bounds} presents upper bounds on information cost and maximal leakage based on Shannon entropy as well as min-entropy in differentially private mechanisms.
In Section \ref{sec:source_coding}, we discuss the connections between differential privacy and source-coding theory, in addition to an exemplary result on adversarial classification in differentially private mechanisms from a rate-distortion perspective. To conclude, in Section \ref{sec:future}, we point out possible research directions on information-theoretic approaches to differential privacy for future work.
\section{Preliminaries \label{sec:preliminaries}}

This section is reserved for a review of some important preliminaries from the differential privacy literature. 
We begin with defining the notion of neighborhood of datasets and the sensitivity of differential privacy. 
\begin{definition}\label{eq:distance}
Two datasets $x$ and $\tilde{x}$ are called neighbors, if the following equality holds
%Hamming (or $l_1$) distance between two datasets $x$ and $\tilde{x}$ is defined by %\cite{r5} as
\begin{equation} \label{eq:hamming}
d(x, \tilde{x})=1 
\end{equation} where $d(.,.)$ denotes the Hamming or $l_1$ distance between the datasets \cite{DR05}.
\end{definition}  

\begin{wrapfigure}{l}{0.48\textwidth}		
  \begin{center}
    \includegraphics[width=0.48\textwidth]{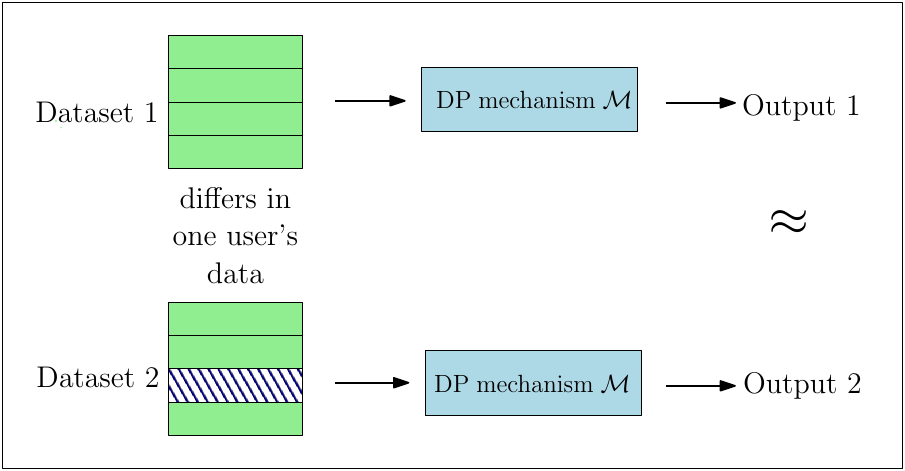}
  \end{center}
  \caption{Symmetric neighborhood of differential privacy 	\label{fig:dp_def}}
\end{wrapfigure} Definition \ref{eq:distance} considers symmetry among neighbors in terms of the size of the dataset as depicted in Figure \ref{fig:dp_def}. This is further relaxed to include the datasets, where neighborhood is due to addition or removal of a record as shown in Figure \ref{fig:dp_def2}. In both cases, neighbors differ in a single row. 
\begin{definition} \label{def:sens}
Global sensitivity, denoted by $s$, of a function (or a query) $q: D \rightarrow \mathbb{R}^k$ is the smallest possible upper bound on the distance between the images of $q$ when applied to two neighboring datasets $x$ and $\tilde{x}$. This means that the $l_1$ distance is bounded as follows $\|q(x)-q(\tilde{x})\|_1 \leq s$ \cite{DMNA06} .
\end{definition} 

Basically, sensitivity of a differentially private mechanism of Definition \ref{def:sens} is the tightest upper bound on the images of a query (a mapping function) for neighbors. It is a function of the type of the query having an opposite relationship with the privacy, since higher sensitivity of the query refers to a stronger requirement for privacy guarantee, consequently more noise is needed to achieve that guarantee.

The original definition of differential privacy makes use of this notion of neighborhood between datasets. An informal definition is depicted in Figure \ref{fig:dp_def}. Accordingly, a mechanism $\mathcal{M}$ is said to be differentially private if for any two neighboring datasets, corresponding outputs of the mechanism, Outputs 1 and 2, are indistinguishable. In other words, the output of a differentially private mechanism is expected to behave in the same way whether or not one contributes the dataset with their data. The following formal definition of differential privacy introduced and studied by Dwork \textit{et al.} in various publications \cite{D06, D08, DR05} clarifies the mathematical meaning of indistinguishability of the outputs corresponding to neighboring datasets. 
%\begin{figure}[h!]
 %\centering
%%\includegraphics[width=1\linewidth]{defdp.eps}
%\includegraphics[width=.45\linewidth]{plots/dp_def.eps}
%\label{fig:dp_def}
 %\end{figure}

%
%\begin{figure}
%\centering
%\begin{minipage}{.35\textwidth}
  %\centering
  %\includegraphics[width=1\linewidth]{plots/dp_def.eps}
  %%\captionof{figure}{A figure}
  %\label{fig:dp_def}
%\end{minipage}%
%\begin{minipage}{.65\textwidth}
  %\centering
  %\includegraphics[width=1\linewidth]{plots/dp_def2.jpg}
  %%\captionof{figure}{Another figure}
  %%\label{}
%\end{minipage}
%\end{figure}

\begin{definition}\label{def:dp}
$(\epsilon, \delta)-$ differential privacy:
A randomized algorithm $\mathcal{M}$ is $(\epsilon, \delta)-$ differentially private if $\forall S \subseteq Range(\mathcal{M})$ and $\forall x, \tilde{x}$ that are neighbors within the domain of $\mathcal{M}$, the following inequality holds.
\begin{equation}\label{ineq:dp}
\Pr\left[\mathcal{M}(x) \in S\right] \leq \Pr\left[\mathcal{M}(\tilde{x}) \in S \right] \mathrm{e}^{\epsilon}+\delta 
\end{equation}
\end{definition} 

For two different privacy measures $\epsilon_1-$DP and $\epsilon_2-$DP where $\epsilon_1, \epsilon_2 >0$, $\epsilon_1-\mathrm{DP} \succeq \epsilon_2-\mathrm{DP}$ denotes that $\epsilon_1-$ DP is a stronger privacy metric than $\epsilon_2-$DP.
Analogous to Definition \ref{def:dp}, there are two other cases of differential privacy where either of the privacy parameters, $\epsilon$ or $\delta$, equals to zero. The ordering of these three cases from the strongest to the weakest privacy metric is as follows
\begin{equation}
\epsilon-\mathrm{DP} \succeq (\epsilon, \delta)-\mathrm{DP} \succeq \delta-\mathrm{DP}. 
\end{equation} 

Dwork's original definition of differential privacy in Definition \ref{def:dp} emanates from a notion of statistical indistinguishability of two different probability distributions given by the next definition.
\begin{definition}[Statistical Closeness] \label{def:stat_close}
Two probability distributions $P_1$ and $P_2$ are said to be $(\epsilon, \delta)-$ close denoted by $P_1 \overset{(\epsilon, \delta)}{\approx}P_2$ over the measurable space $(\Omega, \mathcal{F})$ iff the following inequalities hold.
\begin{align}
P_1(A) \leq \mathrm{e}^{\epsilon} P_2(A) +\delta, \;\; \forall A \in \mathcal{F} \\
P_2(A) \leq \mathrm{e}^{\epsilon} P_1(A) +\delta, \;\; \forall A \in \mathcal{F} 
\end{align}
\end{definition}
Some important properties of statistical closeness are reminded here which will be used in Section \ref{sec:shannon_inf} to prove equality between mutual information functional and differential privacy.
\begin{enumerate}
\item Property 1: Statistical closeness have the following relation with Kullback-Leibler divergence.
\begin{equation}%\label{property1}
P_1 \overset{(\epsilon, 0)}{ \approx }  P_2 \Longrightarrow 
    \begin{gathered}  
D(P_1||P_2) \leq \min \{\epsilon, \epsilon^2 \} \\
D(P_2||P_1) \leq \min \{\epsilon, \epsilon^2 \} 
    \end{gathered} 
\end{equation} Note that, the right hand sides of the inequalities are given in nats.
\item Property 2: Due to Pinsker's inequality, we also have
\begin{equation}
D(P_1||P_2)\leq \epsilon\; \mathrm{nats} \Longrightarrow P_1\overset{(0,\sqrt{\epsilon/2})}{ \approx } P_2. 
\end{equation}
\item Property 3: For any $\epsilon' < \epsilon$ and $\delta'=1-\frac{(\mathrm{e}^{\epsilon'}+1)(1-\delta)}{\mathrm{e}^{\epsilon}+1}$, we have the following relation.
\begin{equation}
P_1 \overset{(\epsilon, \delta)}{ \approx}  P_2 \Longrightarrow P_1 \overset{(\epsilon', \delta')}{ \approx} P_2 
\end{equation} 
\end{enumerate} 

\subsection{How to obtain $\texorpdfstring{\epsilon-}{epsilon}$ and $(\texorpdfstring{\epsilon, \delta}{epsilon, delta})-$ differential privacy?}

A differentially private mechanism is named after the probability distribution of the perturbation applied onto the query output, in the global setting. In the following, we remind the reader of the Laplace distribution and introduce Laplace and Gaussian mechanisms. 
The Laplace distribution, also known as the double exponential distribution, with location parameter $\mu$ and scale parameter $b$ is defined by
\begin{equation}
Lap(x; \mu, b)= \frac{1}{2b} \mathrm{e}^{-\frac{|x-\mu|}{b}}
\end{equation} where its mean equals its location parameter $\mu$ and its variance is $2b^2$.
\begin{definition}
Laplace mechanism \cite{DMNA06} for a function (or a query) $q: D \rightarrow \mathbb{R}^k$ is defined by
\begin{equation}
\mathcal{M}(x, q(.), \epsilon)= q(x)+(Z_1, \cdots, Z_k)
\end{equation} where $Z_i \sim Lap(b=s/\epsilon)$, $i=1,\cdots, k$ denote i.i.d. Laplace random variables.
\end{definition} 

\begin{definition}\label{def:gauss}
Gaussian mechanism \cite{DMNA06} is defined for a function (or a query) $q: D \rightarrow \mathbb{R}^k$ as follows
\begin{equation}
\mathcal{M}(x, q(.), \epsilon, \delta)= q(x)+(Z_1, \cdots, Z_k)
\end{equation} where $Z_i \sim \mathcal{N}(0, \sigma^2)$, $i=1,\cdots, k$ denote i.i.d. Gaussian random variables with the variance $\sigma^2= \frac{2 s^2 \log (1.25/\delta)}{\epsilon^2}$.
\end{definition}
\begin{theorem}{\cite{DR05}} For any $\epsilon, \delta \in(0,1)$, the Gaussian mechanism satisfies $(\varepsilon,\delta)$-differential privacy.
\end{theorem}
\begin{remark}As an alternative to Laplacian perturbation applied on the query output which results in $(\epsilon, 0)-$ differential privacy, Gaussian noise provides a more relaxed privacy guarantee, that is $(\epsilon, \delta)-$ differential privacy. However, in some cases, application of Gaussian noise becomes more useful. Vector valued Laplace mechanisms require the use of $l_1$- sensitivity whereas the vector-valued Gaussian mechanism allows $l_1$ or $l_2$ sensitivity, where $l_2$ sensitivity is defined as $\max_{x, \tilde{x}}\|q(x)-q(\tilde{x})\|_{2} \leq s$, for neighboring $x$ and $\tilde{x}$. Dependent on the query function, when the $l_2$ sensitivity is significantly lower than $l_1$ sensitivity, Gaussian mechanism requires much less noise. 
\end{remark}
\begin{remark}[The optimal $ \epsilon-$ differentially private mechanism]
%\paragraph*{The optimal $(\epsilon,0)-$ differentially private mechanism}
A natural question that comes to mind is if we can do better than the Laplace mechanism. The work in \cite{GV14} improves the Laplace mechanism of \cite{DMNA06} by characterizing the fundamental trade-off between the differentially private mechanism's privacy and utility to define an \textit{optimal} $\epsilon-$ mechanism. Accordingly, \cite[Theorem 1]{GV14} shows that such a mechanism is obtained by applying a staircase-shaped probability distribution as the perturbation on real and integer-valued query functions in the low-privacy regime (i.e. when $\epsilon$ is large). Laplace mechanism outperforms the optimal $(\epsilon, 0)-$ mechanism in the high privacy regime. 
\end{remark}
\section{Shannon Information and Relative Entropy as a Privacy Constraint \label{sec:shannon_inf}}

This first main part of the tutorial is dedicated for presentation of information-theoretic quantities adapted to be used as privacy constraint in systems that deploy $(\epsilon, \delta)-$ differential privacy.
%\subsection{Definitions}
\begin{definition}[$\epsilon-$Mutual-Information differential privacy (MI-DP) \cite{CY16}] \label{def:MI-dP}
For a dataset $X^n=(X_1, \cdots, X_n)$ with the corresponding ML output $Y$ according to the randomized mechanism represented by $\mathcal{M}=P_{Y|X^n}$, \textit{mutual information differential privacy} (MI-DP) is defined as
\begin{equation} \label{eq:MI-DP}
\underset{i, P_{X^n}}{\sup} I(X_i;Y|X^{-i}) \leq \epsilon \;\mathrm{nats}
\end{equation} where $X^{-i}=\{X_1,\cdots,X_{i-1},X_{i+1},\cdots, X_n\}$ denotes the dataset entries excluding $X_i$.
\end{definition}
$\epsilon-$MI-DP definition of Cuff \textit{et al.} in \cite{CY16} combines the Shannon information with the notion of \textit{identifiability} which is defined using the Bayesian approach on indistinguishability of the neighboring datasets. Accordingly, a mechanism $\mathcal{M}$ satisfies  \textbf{$\epsilon-$identifiability} for some positive and real $\epsilon$ if the following inequality holds for any neighboring entries $x, \tilde{x} \in \mathcal{D}^n$ and any output $y \in \mathcal{D}^n$.
\begin{equation}\label{ineq:identifiable}
P_{X|Y}(x|y)\leq \mathrm{e}^{\epsilon} P_{X|Y}(\tilde{x}|y) 
\end{equation} Both $\epsilon-$MI-DP and $\epsilon-$identifiability are subject to the implicit strong adversary assumption \cite{CY16} (also called as the \textit{informed adversary} in \cite{DMNA06}) where the adversary has the knowledge of all but a single entry in a dataset and aims to discover the last one. 
The condition in (\ref{ineq:identifiable}) suggests that for small values of $\epsilon$, neighboring datasets are indistinguishable based on the posterior probabilities of the output. This is what makes it hard to associate the representation of the data and the data owner, which translates to re-identification. Another line of work in \cite{WYZ16} defines the mutual information-based differential privacy as a lossy source coding problem without the maximization taken over all possible dataset distributions. Definition \ref{def:MI-dP} differs from the information theoretic definitions of original differential privacy by incorporating that no assumptions are made on prior dataset distributions. Maximization over all possible input distributions in (\ref{eq:MI-DP}) assures that the differential privacy is a property of the mechanism resembling the well-known formula of the Shannon capacity.

Next, we remind the reader of the so-called \textit{Kullback-Leibler differential privacy}.
\begin{definition}[$\epsilon-$Kullback-Leibler (KL) differential privacy \cite{CY16}] \label{def:KL-DP}
A randomized mechanism $P_{Y|X}$ guarantees $\epsilon-$ KL-differential privacy, if the following inequality holds for all its neighboring datasets $x$ and $\tilde{x}$, 
\begin{equation} \label{eq:KLdp}
D(P_{Y|X=x}||P_{Y|X=\tilde{x}}) \leq \mathrm{e}^{\epsilon}.
\end{equation}
\end{definition}
\subsection{Main Result}
Using information-theoretic quantities to study privacy may not be a brand new approach, nonetheless, the following result draws the strongest link between the two areas. Ordering and equivalence of $\epsilon-$MI-DP and differential privacy is given by Theorem \ref{theo:MI-DP}.
\begin{theorem}[\cite{CY16}] \label{theo:MI-DP}
The following chain of inequalities hold
\begin{equation} \label{ineq1}
\epsilon-\textrm{DP} \succeq \epsilon-\textrm{MI-DP}\succeq (\epsilon,\delta)-\textrm{DP}
\end{equation} 
Conditioned on the cardinality of the input $\mathcal{X}_i$ or the output $\mathcal{Y}$ of the differentially private mechanisms, an equivalence is achieved between $\epsilon-$MI-DP and $(\epsilon, \delta)-$ differential privacy. Then, we have
\begin{equation}
\epsilon-\textrm{MI-DP}= (\epsilon,\delta)-\textrm{DP}
\end{equation} 
The case $(\epsilon,\delta)-\textrm{DP}\succeq \epsilon-\textrm{MI-DP}$ depends on the cardinality bound $\min \left\{|\mathcal{Y}|, \underset{i}{\max} |\mathcal{X}_i|\right\}$.
\end{theorem} 
\paragraph*{The sketch of the proof of Theorem \ref{theo:MI-DP} \cite{CY16}}
As is well-known, (un/conditional) mutual information can be represented as a function of relative entropy. The proof of Theorem \ref{theo:MI-DP} starts off by proving an even more powerful chain of in/equalities among all three variations $\epsilon$, $(\epsilon, \delta)$ and $\delta-$ differential privacy, mutual information differential privacy and the Kullback-Leibler differential privacy.
The chain of inequalities in (\ref{ineq1}) is expanded out as follows.
\begin{equation} \label{ineq:ordering}
\epsilon-\mathrm{DP} \overset{(a)}{\succeq} \mathrm{KL-DP} \overset{(b)}{\succeq} \epsilon-\mathrm{MI-DP}\overset{(c)}{\succeq} \delta-\mathrm{DP} \overset{(d)}{=} (\epsilon, \delta)-\mathrm{DP}
\end{equation} %The ordering in (\ref{ineq:ordering}) are proven by three main rules regarding the relationship of statistical closeness in Definition \ref{def:stat_close} with Kullback-Leibler divergence and their corresponding results as follows. 
(\ref{ineq:ordering}) shows that an $\epsilon-$DP mechanism also guarantees $\epsilon-$MI-DP. 
Relations (a) and (b) in (\ref{ineq:ordering}) are results of Property 1 of statistical closeness given by Definition \ref{def:stat_close}. Ordering in (b) is achieved as follows
\begin{align}
 D(P_{Y|X^n=x^n}||P_{Y|X^{-i}=x^{-i}}) 
&=D\left(P_{Y|X^n=x^n}||\mathbb{E}\left[P_{Y|X_i=\tilde{X},X^{-i}=x^{-i}} \right]\right) \label{eq:step1}\\
&\leq\mathbb{E}\left[D(P_{Y|X^n=x^n}||P_{Y|X^{-i}=\tilde{X}, X^{-i}=x^{-i}})\right] \label{eq:step2}\\
& \leq \epsilon \; \mathrm{nats}  \label{eq:step3}
\end{align} for $\tilde{X} \sim P_{X_i|X^{-i}=x^{-i}}$ and $x^{-i}$ denotes an instance of $X^{-i}$. Thus, in (\ref{eq:step1}) we use $P_{Y|X^{-i}=x^{-i}}=\mathbb{E}\left[P_{Y|X_i=\tilde{X},X^{-i}=x^{-i}}\right]$.
The steps in (\ref{eq:step2}) and (\ref{eq:step3}), respectively follow due to Jensen's inequality and the definition of mutual information based on relative entropy, that is
\begin{equation}
I(X_i;Y|X^{-i})=\mathbb{E}\left[D(P_{Y|X^n=\bar{X}^n}||P_{Y|X^{-i}=\bar{X}^{-i}})\right] 
\end{equation} where $\bar{X}^n \sim P_{X^n}$. Ordering (c) that states $\epsilon-\mathrm{MI-DP}\succeq \delta-\mathrm{DP}$ is a consequence of Lemma \ref{lemma2}.
\begin{lemma}[\cite{CY16}] \label{lemma2}
The following statement is satisfied with respect to the relation between $\epsilon-$MI-DP and $(\delta)-$ DP.
\begin{equation}\label{eq:orderingc}
\epsilon-\mathrm{MI-DP} \Longrightarrow (0, \sqrt{2\epsilon})-\mathrm{DP}
\end{equation} (\ref{eq:orderingc}) is tightened as $\epsilon-\mathrm{MI-DP} \Longrightarrow (0, \delta')-\mathrm{DP}$ for $\epsilon \in [0,\ln2]$ for $\delta'=1-2h^{-1}(\ln2-\epsilon)$ and $h^{-1}$ denotes the inverse of the binary entropy function.
\end{lemma}
Lastly, ordering (d) in (\ref{ineq:ordering}) is due to Property 3 given by Definition \ref{def:stat_close}. 
The reader is referred to \cite[Section 3.3.]{CY16} for the full proof.
\begin{remark}
The major strength of $\epsilon-$MI-DP over other alternative mutual information based definitions of differential privacy lies in \textbf{the maximization taken over all possible input distributions to capture the fact that differential privacy does not require a particular distribution of the input.} %Moreover,  which was transmitted implicitly by using the notion of neighboring databases. 
Moreover, from a stochastic perspective, conditional mutual information reflects the strong adversary assumption of differential privacy and establishes another major strength of $\epsilon-$MI-DP that is based on Dwork's standard definition of differential privacy which originally stems from this assumption. Conditioning on the remaining entries of the dataset in $\epsilon-$MI-DP demonstrates that the adversary has the knowledge of the entire dataset except for one entry, which was transmitted implicitly by using the notion of neighboring datasets in the original stochastic definition of differential privacy. From a practical point of view, another major strength of Definition \ref{def:MI-dP} lies in the ability to transfer information-theoretic rules and properties defined for Shannon information and related measures onto differential privacy.
\end{remark}
 Next part provides some of the well-known information-theoretic rules that also apply to differential privacy as a consequence of MI-DP and the ordering in (\ref{ineq:ordering}).
%%%%%%%%%%%%%%%%%%%%%%%%EDITED AS OF HERE
\subsection{Composability of $\epsilon-$MI-DP via information-theoretic rules\label{subsec:properties}}
This part is dedicated for some of the well-known properties of mutual information which are now directly applicable on $\epsilon-$MI-DP. 
\begin{enumerate}
\item Bounding the conditional mutual information: If $X$ is independent of $Z$, then the following inequality holds.
\begin{equation}
I(X;Y|Z) \geq I(X;Y)  
\end{equation}
\item Consequence of data processing inequality: If $X\rightarrow Y \rightarrow Z$ form a Markov chain in that order that is $X$ and $Z$ are conditionally independent given $Y$, then the following inequality holds.
\begin{equation}
I(X;Y|Z) \leq I(X;Y)
\end{equation}
\item Chain rule:
\begin{equation} 
\!\!\! I(X;Y,Z)=I(X;Z)+I(X;Y|Z) 
\end{equation} 
\item Independence: If the differentially private mechanism $\mathcal{M}=P_{Y|X^n}$ satisfies $\epsilon-$MI-DP where $\{X_i\}_{i=1}^{n}$ are mutually independent, then the following chain of inequalities hold.
\begin{equation}
\!\!\!\underset{i, P_{X^n}}{\sup} I(X_i;Y) \leq \underset{i, P_{X^n}}{\sup} I(X_i;Y|X^{-i})\leq \epsilon 
\end{equation}
\end{enumerate}
Some of the fundamental rules of mutual information enlisted above are transferred onto differential privacy as a result of Theorem \ref{theo:MI-DP}. Several important properties of differential privacy are straightforward to prove in this mutual information based approach. Next, we prove the composition theorem of differential privacy with the aid of these properties. Originally, composability -an important property of $(\epsilon, 0)-$DP- states that a number of queries under differential privacy also collectively satisfies differential privacy where the privacy budget of the collection is scaled proportionally to the number of queries \cite{DRV10, KOV15}. Corollary \ref{composition_corr} is a reflection of the composition theorem for $(\epsilon, 0)-$ differential privacy onto $\epsilon-$MI-DP, which shows that the composability can be defined and proven using information-theoretic quantities and their corresponding properties.  
\begin{corollary}[Composition of $\epsilon-$MI-DP \cite{CY16}] \label{composition_corr}
For randomized mechanisms $\mathcal{M}_j=P_{Y_j|X^n}$ that individually satisfy $\epsilon-$MI-DP with $k$ conditionally independent outputs $\{Y_1, \cdots, Y_k\}$ given the input $\{X_1, \cdots, X_n\}$, the collection of $k$ mechanisms $\mathcal{M}_k=P_{Y_k|X^n}$ also satisfies $\epsilon-$MI-DP with the privacy parameter $\sum_j^k \epsilon_j$.
\end{corollary}
\proof{ For any $P_{X^n}$ and $i$, the collection of $P_{Y_k|X^n}$ satisfies $\epsilon-$MI-DP which is bounded as follows:
\begin{align}
I(X_i;Y|X^{-i})&= \sum_{l=1}^m I(X_i;Y_l|X^{-i}, Y^{l-1}) \label{1stline} \\
&\leq \sum_{l=1}^m I(X_i;Y_l|X^{-i}) \label{2ndline}
\end{align}
(\ref{1stline}) follows due to the chain rule given by Property 3 in Section \ref{subsec:properties}. Step in (\ref{2ndline}) uses a property of the data-processing inequality (Property 2 in Section \ref{subsec:properties}) due to the conditional independence between $X_i$ and $Y^{l-1}$ given $Y_l$.  
Finally, (\ref{3rdline}) substitutes Definition \ref{def:MI-dP} as given below.
\begin{equation}
I(X_i;Y|X^{-i})\leq \sum_{l=1}^{m} \epsilon_j \; \textrm{nats} \label{3rdline}
\end{equation} 
}
%\end{proof}
This result completes the first main part of the tutorial.% We have shown that owing to \cite{CY16}, 

\section{Information-Theoretic Bounds on Differential Privacy \label{sec:bounds}}

In this section, we review four selective publications \cite{MMP+10,CF12,BK11,AA11} that present upper bounds on the performance of differentially private mechanisms using different metrics. We begin with the two-party differential privacy in the distributed setting in the upcoming part.

\subsection{Bounding the Information Cost}
Contrarily to the common client-server setting where the server answers queries of clients based on its access policy, in the two-party distributed setting parties execute their analysis on joint data where the aim is to provide a two-sided privacy guarantee for each party's data. In such a setting, each side sees the protocol/mechanism as a differentially private version of the other side's input data. Information cost of a two-user differential privacy model in such a setting refers to the amount of information gathered from each party's inputs using the exchanged messages. In order to prove the usefullness and practicality of differential privacy, McGregor \textit{et al.} characterizes in \cite{MMP+10} a fundamental connection between the information cost and differential privacy. Accordingly, the authors presents an upper bound on the information cost of such a mechanism by defining the cost as the mutual information between the inputs and the random transcript of the mechanism denoted $\Pi(.,.)$ which simply is the sequences of exchanged messages between the two parties.
\begin{definition}[Information Cost]
For two inputs $X$ and $Y$ of a two-party mechanism $\mathcal{M}$ with probability distribution $P$, the information cost of the mechanism is defined as
\begin{equation}
{I}cost_{P}(\mathcal{M})=I(X,Y;\Pi(X,Y))
\end{equation}
\end{definition}
For a finite alphabet $\Sigma$, the two-party $\epsilon-$ differential privacy mechanism $\mathcal{M}(x,y)$ with $x,y \in \Sigma^n$ and every distribution $P$ defined on $\Sigma^n \times \Sigma^n$, the information cost of this mechanism satisfies the upper bound 
\begin{equation} \label{Icost_upper}
{Icost}_{P}(\mathcal{M}) \leq 3 \epsilon n.
\end{equation}
For the special case of $\Sigma=\{0,1\}$ and $P$ is the uniform distribution, the bound in (\ref{Icost_upper}) is improved to $1.5 \epsilon^2 n$ \cite[Proposition 4.3]{MMP+10}.

\paragraph*{Derivation of the upper bounds}
For the two-party random input denoted by $T=(X_1,\cdots, X_n, Y_1, \cdots, Y_n)$ and independent sample $T'$ from the uniform distribution $P$, we have
\begin{align}
I(\Pi(T);T)& = H(\Pi)-H(\Pi|T) \notag \\
&=\mathbb{E}_{(t,\pi) \leftarrow (T,\Pi(T))} \log \frac{\Pr\left[\Pi[T]=\pi|T=t\right]}{\Pr\left[\Pi[T]=\pi \right]} \\
&\leq 2 (\log_2 \mathrm{e}) \epsilon n \label{infcost_bound}
\end{align} (\ref{infcost_bound}) is equivalent to the right hand side of (\ref{Icost_upper}) and obtained using the following interval for any $t$ and $t'$.
\begin{equation}
\mathrm{e}^{(-2\epsilon n)}\leq \frac{\Pr\left[\Pi(t)=\pi\right]}{\Pr\left[\Pi(t')=\pi\right]} \leq \mathrm{e}^{(2\epsilon n)}
\end{equation}
The improvement is achieved by setting $\Sigma= \{0,1\}$ for a uniform distribution $P$ as follows.
\begin{align}
I(T;\Pi(T))&= \sum_{i\in [2n]} I(T_i; \Pi(T)|T_1\cdots T_{i-1}) \label{eq:step_i}\\
&=\sum_{i\in [2n]}H(T_i|T_1\cdots T_{i-1})-H(T_i|\Pi(T) T_1 \cdots T_{i-1})  \label{eq:step_ii} \\
& \leq \sum_{i\in [2n]} (1- H(\mathrm{e}^{\epsilon}/2)) \label{eq:step_iii}\\
& \leq  \sum_{i\in [2n]}  \frac{\epsilon^2}{2 \ln 2} \label{eq:step_iv}
%&\overset{(iv)}{=} \log_2 (\mathrm{e}) \epsilon^2 n
\end{align}
The first term in (\ref{eq:step_ii}) equals 1 since each $T_i$ is independent and uniform in $P$. Due to the differential privacy property and the Bayes rule, we have $\forall t_1, \cdots, t_{i-1}, \pi$ the ratio confined in the interval $(\mathrm{e}^{-\epsilon}, \mathrm{e}^{\epsilon})$ as given by
\begin{equation}
\mathrm{e}^{-\epsilon} \leq \frac{\Pr\left[T_i=0|T_1, \cdots, T_{i-1}=t_1,\cdots, t_{i-1}, \Pi[T]=\pi \right]}{\Pr\left[T_i=1|T_1, \cdots, T_{i-1}=t_1,\cdots, t_{i-1}, \Pi[T]=\pi \right]} \leq \mathrm{e}^{\epsilon} \label{int_bayes}
\end{equation}
Accordingly, the second term in (\ref{eq:step_ii}) is bounded by the entropy in (\ref{eq:step_iii}). Finally, in (\ref{eq:step_iv}), the base of the logarithm is changed and summed over $2n$ terms to get $\log_2 (\mathrm{e}) \epsilon^2 n$.
\cite{D12} presents an adaptation of the upper bound in (\ref{Icost_upper}) to the mutual information between the distribution over the inputs of an $\epsilon-$ differentially private mechanism and the mechanism's output by replacing the second party's input with a constant to obtain the same behavior of $3\epsilon n$. Accordingly, for a query $q:(\mathbb{Z}^+)^d \rightarrow \mathbb{R}^k$, an $\epsilon-$ differentially private mechanism $\mathcal{M}: (\mathbb{Z}^+)^d \rightarrow P \mathbb{R}^k$ and a dataset size of $n$, the mutual information $I(X;\mathcal{M}(X))$ is upper bounded by $3 \epsilon n $. Bounding the size of the dataset by $n$, allows the input distribution to be narrowed down to $X \in [n]^d$ for $[n]=\{0,1,\cdots,n\}$. This results in the direct application of the upper bound (\ref{Icost_upper}) by McGregor \textit{et al.} when the second party's input is set to be a constant.
\begin{remark}
(\ref{Icost_upper}) bounds the information cost as a function of the privacy budget of a differentially privacy mechanism and combined with \cite{BB+10}, the result signifies that any mechanism that satisfies differential privacy can be compressed. Additionally, well-known bounds for the information cost in various settings can be employed to characterize the gap between the optimal and computational differential privacy mechanisms.
\end{remark}
%%%%%%%%%%%%%%%%%%%%%%%%%%%%%%%%%%%%%%%%%%%%%%%%%%%%%%%%%%%%%%%%%%%%%%%%%%%%%%%%%%%%%%%%%%%%%%%%%%%%%%%%%%%%%%%%%%%%%%%%%%%%%%%%%%%%%%%%%%%%%%%%%%%%%%%%%%%%%%%%%%%%%%%%%%%%%%%%%%%%%%%%%%%%%%%%%%%%%%%
%%%%%%%%%%%%%%%%%%%%%%%%%%%%%%%%%%%%%%%%%%%%%%%%%%%%%%%%%%%%%%%%%%%%%%%%%%%%%%%%%%%%%%%%%%%%%%%%%%%%%%%%%%%%%%%%%%%%%%%%%%%%%%%%%%%%%%%%%%%%%%%%%%%%%%%%%%%%%%%%%%%%%%%%%%%%%%%%%%%%%%%%%%%%%%%%%%%%%%%
%%%%%%%%%%%%%%%%%%%%%%%%%%%%%%%%%%%%%%%%%%%%%%%%%%%%%%%%%%%%%%%%%%%%%%%%%%%%%%%%%%%%%%%%%%%%%%%%%%%%%%%%%%%%%%%%%%%%%%%%%%%%%%%%%%%%%%%%%%%%%%%%%%%%%%%%%%%%%%%%%%%%%%%%%%%%%%%%%%%%%%%%%%%%%%%%%%%%%%%
\subsection{Upper bound on maximal leakage}  
\cite{CF12} is one of the first examples of the line of work that modeled the problem of defining the optimal mapping of the input data to a privatized output in order to determine the privacy-utility trade-off by using rate-distortion theory. Additionally, the authors compare differential privacy with the maximum information leakage to prove that differential privacy does not grant privacy with regard to average and maximal leakage. Their model is designed as a noiseless communication channel between two parties to transmit a number of measurements denoted $Y \in \mathcal{Y}$ to the receiving end, as well as a set of variables $X\in \mathcal{X}$ which is required to remain private to the sender. $X$ and $Y$ follow the joint distribution $(Y,X) \sim p_{Y,X}(y,x), \; (y,x)\in \mathcal{Y} \times \mathcal{X}$. 

$\epsilon-$ \textit{information privacy} is defined as follows in the sense of a differentially private mechanism as a stronger alternative to the Dwork's original definition. Accordingly,  $\epsilon-$information privacy captures the fundamental aim of privacy of resisting to notable change in the conditional prior and posterior probabilities of the features given the output.
\begin{definition}[\cite{EGS3}] \label{def:epsilon_inf}
A privacy preserving mapping defined by the transition probability $p_{Y|\mathbf{X}}(.|.)$ for a set of features $\mathbf{X}=(X_1, \cdots, X_n)$ where $X_i\in \mathcal{X}, \; y\in \mathcal{Y}$ provides $\epsilon-$ differential privacy
\begin{equation}
\mathrm{e}^{-\epsilon} \leq \frac{p_{\mathbf{X}|Y}(\mathbf{x}|y)}{p_{\mathbf{X}}(\mathbf{x})} \leq \mathrm{e}^{\epsilon}
\end{equation}  for all $y \in \mathcal{Y}: p_Y(y)>0$ if $\forall x \subseteq \mathcal{X}^n$.
\end{definition}

Definition \ref{def:epsilon_inf} is used for bounding the \textit{maximal (information) leakage} defined by  
\begin{equation} \label{max_leak}
\max_{y \in \mathcal{Y}} \; H(X) - H(X|Y=y).
\end{equation} Maximal leakage refers to the maximum cost gain achieved by the adversary using a single output. The main result of \cite{CF12} connecting $\epsilon-$ information privacy to differential privacy is given by the next theorem.
\begin{theorem}[Upper bound on maximal leakage of differential privacy\cite{{CF12}}]
If a privacy-preserving mapping $p_{Y|\mathbf{X}}(.|.)$ is $\epsilon-$ information private for some $supp(p_Y)=\mathcal{Y}$ then it provides at least $2\epsilon-$ differential privacy and the maximal leakage is at most $\frac{\epsilon}{\ln 2}$.
\end{theorem}
\proof{
For neighbors $\mathbf{x}_1$ and $\mathbf{x}_2$, we have for $p_{Y|\mathbf{X}}(.|.)$ and a subset $B \subseteq \mathcal{Y}$
\begin{align}
\frac{\Pr[Y \in B|\mathbf{X}=\mathbf{x}_1]}{\Pr[Y \in B|\mathbf{X}=\mathbf{x}_2]}
&=\frac{\Pr[\mathbf{X}=\mathbf{x}_1| Y \in B]\Pr[\mathbf{X}=\mathbf{x}_2]}{\Pr[\mathbf{X}=\mathbf{x}_2| Y \in B]\Pr[\mathbf{X}=\mathbf{x}_1]}\\
&\leq  \mathrm{e}^{2 \epsilon} \label{eq:bound_min}
\end{align} Bounding step in (\ref{eq:bound_min}) is a result of Definition \ref{def:dp}. The maximum amount of information that is leaked from $\epsilon-$ information private mapping (\ref{max_leak}) is bounded as given below.
\begin{align}
H(X)- H(X|Y=y)
& = \underset{\mathbf{x}\in \mathcal{X}^n}{\sum}p_{\mathbf{X}|Y}(\mathbf{x}|y) p_Y(y) \log \left( \frac{p_{\mathbf{X}|Y}(\mathbf{x}|y)}{p_{\mathbf{X}}(\mathbf{x})}\right)\\
&\overset{(i)}{\leq} \underset{\mathbf{x}\in \mathcal{X}^n, y \in \mathcal{Y}}{\sum}p_{\mathbf{X}|Y}(\mathbf{x}|y) p_Y(y)\log \mathrm{e}^{\epsilon} \\
& \overset{(ii)}{=} \frac{\ln \mathrm{e}^{\epsilon}}{\ln 2} 
\end{align} Step (i) results from applying the upper bound of Definition \ref{def:epsilon_inf} and from changing the range of the sum. In step (ii), the base of the logarithmic function is changed and the summation equals to 1, thus we get $\epsilon/ \ln 2$.}

%%%%%%%%%%%%%%%%%%%%%%%%%%%%%%%%%%%%%%%%%%%%%%%%%%%%%%%%%%%%%%%%%%%%%%%%%%%%%%%%%%%%%%%%%%%%%%%%%%%%%%%%%%%%%%%%%%%%%%%%%%%%%%%%%%%%%%%%%%%%%%%%%%%%%%%%%%%%%%%%%%%%%%%%%%%%%%%%%%%%%%%%%%%%%%%%%%%%%%%
%%%%%%%%%%%%%%%%%%%%%%%%%%%%%%%%%%%%%%%%%%%%%%%%%%%%%%%%%%%%%%%%%%%%%%%%%%%%%%%%%%%%%%%%%%%%%%%%%%%%%%%%%%%%%%%%%%%%%%%%%%%%%%%%%%%%%%%%%%%%%%%%%%%%%%%%%%%%%%%%%%%%%%%%%%%%%%%%%%%%%%%%%%%%%%%%%%%%%%%
%%%%%%%%%%%%%%%%%%%%%%%%%%%%%%%%%%%%%%%%%%%%%%%%%%%%%%%%%%%%%%%%%%%%%%%%%%%%%%%%%%%%%%%%%%%%%%%%%%%%%%%%%%%%%%%%%%%%%%%%%%%%%%%%%%%%%%%%%%%%%%%%%%%%%%%%%%%%%%%%%%%%%%%%%%%%%%%%%%%%%%%%%%%%%%%%%%%%%%%
\subsection{Upper bound on maximal leakage based on min-entropy}
This part presents the review of an upper bound on the maximal leakage of $\epsilon-$ differential privacy by \cite{BK11}. The distinction of the work stems from using \textit{min-entropy} rather than Shannon entropy. The ultimate goal of \cite{BK11} is to compare and formally characterize connections between differential privacy and information-theoretic leakage. The main contribution is establishing such a connection by upper bounding the information leakage in terms of differential privacy as a function of the privacy budget. 

\cite{BK11} justifies the use of min-entropy by its association to strong security guarantees. For $X$ and $Y$, respectively denoting the input and output to a probabilistic program and the conditional distribution, $P_{Y|X}$ is characterized by the program's semantics and composes an information-theoretic channel between $X$ and $Y$. In this setting, the adversary aims to infer the value of $X$ upon reception of the output $Y$. The unconditional min-entropy $H_{\infty}(X)$ of $X$  is defined by
%\addtocounter{equation}{1}
\begin{equation}
H_{\infty}(X)=-\log \underset{x}{\max} P_X(x),
\end{equation} whereas the conditional min-entropy $H_{\infty}(Y|X)$ of $P_{Y|X}$ yields
\begin{equation}
H_{\infty}(Y|X)=-\log \sum_y P_Y(y) \underset{x}{\max} P_{X|Y}(x,y).
\end{equation}
The min-entropy-based leakage denoted by $L$ is the difference between $H_{\infty}(X)$ and $H_{\infty}(Y|X)$ depending on both the channel $P_{Y|X}$ and the input distribution $P_X$. Min-entropy based maximal leakage $ML(P_{Y|X})$ is given by
%\addtocounter{equation}{1}
\begin{equation} \label{def:max_leak}
ML(P_{Y|X})= \underset{P_X}{\max} (H_{\infty}(X)-H_{\infty}(Y|X)). 
\end{equation} For channels of a single bit of range, that is when $Range(X)=Range(Y)=\{0,1\}$, \cite[Theorem 3]{BK11} states that for an $\epsilon-$ differentially private channel $P_{Y|X}$, the maximal leakage is upper bounded by
\begin{equation}\label{ML_Upper}
ML(P_{Y|X}) \leq \log \frac{2 \mathrm{e}^{\epsilon}}{1+\mathrm{e}^{\epsilon}}
\end{equation} The bound in (\ref{ML_Upper}) is proven to apply to channels of arbitrary finite range in \cite[Corollary 1]{BK11}. Accordingly, the channel $P_{Y|X}$ is summarized in Table \ref{table1} for $\sum_i^n p_i = \sum_i^n q_i =1$.
\begin{table}[htp]
 \centering
\caption{The channel $P_{Y|X}$ with $X=\{0,1\}$ and $Y=\{y_1, y_2, \cdots, y_n\}$.}
\label{table1}
\begin{tabular}{|c|c c c|}
\hline
$P_{Y|X}$ & $Y=y_1$ &$\cdots$ & $Y=y_n$ \\ \hline
$X=0$        & $p_1$   & $\cdots$  & $p_n$   \\ %\hline
$X=1$        & $q_1$   & $\cdots$  & $q_n$  \\\hline
\end{tabular}
\end{table} 

For an $\epsilon-$ differentially private channel $P_{\bar{Y}|X}$, where the output $\bar{Y}$ is defined over the range $\{0,1\}$, the leakage of $P_{Y|X}$ and that of $P_{\bar{Y}|X}$ coincide. Similarly, for the channel $P_{\bar{Y}|X}$ we have the following matrix of probabilities for $I=\{i| p_i \leq q_i \}$. 
In Table \ref{table2}, $\bar{p}$ and $\bar{q}$ respectively denote the sums over $I$ as $\sum_{i \notin I} p_i$ and $\sum_{i \in I} q_i$. Hence their respective complements yield $1- \bar{p}=\sum_{i \in I} p_i$ and $1-\bar{q}= \sum_{i \notin I} q_i$. Plugging in \cite[Theorem 3]{BK11} with the definition of min-entropy based maximal leakage, the equivalence of $ML(P_{Y|X})$ and $ML(P_{\bar{Y}|X})$ is proven by
\begin{align}
ML(P_{Y|X})&= \log \sum_y \max_x P_{Y|X}(y,x) \\
&=\log (\bar{p}+\bar{q}) \label{lastline}
\end{align} since (\ref{lastline}) is $ML(P_{\bar{Y}|X})$.

\begin{table}[htp]
 \centering
\caption{The channel $P_{\bar{Y}|X}$ with $X, \bar{Y}=\{0,1\}$.}
\label{table2}
\begin{tabular}{|c|c c|}
\hline
$P_{\bar{Y}|X}$ & $\bar{Y}=0$ & $\bar{Y}=1$ \\ \hline
$X=0$        & $\bar{p}$   & $1-\bar{p}$   \\ %\hline
$X=1$        & $\bar{q}$   & $1-\bar{q}$  \\ \hline
\end{tabular}
\end{table} 
 Additionally, $\epsilon-$differential privacy of the channel $P_{Y|X}$ guarantees that $q_i \leq \mathrm{e}^{\epsilon}$ for every $i\in I$ and thus, $\bar{q} \leq \mathrm{e}^{\epsilon} \bar{p}$. Same applies to $p_i \leq \mathrm{e}^{\epsilon}$ for every $i \notin I$. 

\subsection{Information-Theoretic Post-processing of Differential Privacy \label{subsec:post-proc}}

The post-processing property is one of the important features of differential privacy and ensures that the privacy protection of a differentially private mechanism is not affected by arbitrary computations applied on the mechanism's output \cite{DR05}. In other words, it is impossible to \textit{undo} the privacy guarantee of differential privacy by post-processing the data. More formally, if the mechanism $\mathcal{M}: \mathbb{N}^{|\mathcal{X}|}\rightarrow R $ satisfies $(\epsilon, \delta)-$ differential privacy, for any arbitrary mapping $f: R \rightarrow R'$, $f \circ \mathcal{M}: \mathbb{N}^{|\mathcal{X}|}\rightarrow R'$ also satisfies $(\epsilon, \delta)-$ differential privacy.

A simpler version of the problem of investigating the connection between differential privacy and min-entropy leakage in \cite{BK11} is initiated by \cite{AC10, AA11} for an individual rather than the entire universe of databases. In \cite{AC10, AA11}, the authors
consider a model where information leakage is used to measure the amount of information that an attacker can learn about
the database which also allows to quantify the utility of the query via min-entropy. Applying Bayesian post-processing on the differentially private output of the mechanism, it is shown that the utility function is closely related to conditional min-entropy and to the min-entropy leakage. 
\section{Differential Privacy as a Source-coding Problem \label{sec:source_coding}}

Several works study the connection between differential privacy and (lossy) source-coding from various aspects \cite{ZLV09, WYZ16, M12, PG21, PKS20, CF12} and some tailored the rate-distortion theory to identify a trade-off between privacy and distortion. \cite{CF12} is one of the first examples that model differential privacy using a rate-distortion perspective establishing a trade-off between privacy and utility. The authors set the amount of information obtained by the adversary (i.e. the leakage) as the cost gain and minimize it subject to a set of utility constraints, which reflect the role of the distortion function in the original setting of the rate-distortion theory. On the other hand, in \cite{WYZ16}, the distortion between the input and output of the mechanism is used to determine the number of rows that differ and it is minimized subject to three different privacy metrics, in order to establish how many rows need to be modified to preserve the privacy guarantee. Accordingly, the distortion is defined as the Hamming distance $d$ between the input and output of a dataset as $d: \mathcal{D}^n \times \mathcal{D}^n \rightarrow \mathbb{N}$. The contribution of \cite{WYZ16} is to demonstrate a connection between identifiability, differential privacy and \textit{the mutual-information privacy} (MIP) that is defined by $I(X;Y)$ for the input $X$ and output $Y$.
The privacy-distortion problem of \cite{WYZ16} is defined as follows.
\begin{align}
&\min_{P_{Y|X}}I(X;Y) \\
&\textrm{s.t.}\; \mathbb{E}[d(X,Y)]\leq D,\\
& \sum_{y\in \mathcal{D}^n}p_{Y|X}(y|x)= 1, \forall x \in \mathcal{D}^n, \\
&p_{Y|X}(y|x)\geq 0, \forall x,y \in \mathcal{D}^n 
\end{align}  %This is identical with the original definition of rate-distortion function. 
The main objective of \cite{WYZ16} is to investigate and explain the relation between identifiability, differential privacy and MIP in order to compare them.  
The authors show that there exists a privacy mechanism which minimizes both $I(X;Y)$ and the identifiability. The \textbf{privacy-distortion function} denoted as 
$\epsilon^*(D)$ refers to the smallest differential privacy level for a given maximum allowable distortion $D$. The mutual information based privacy level is bounded as follows
\begin{equation}
\epsilon^*(D) \leq \epsilon \leq \epsilon^*(D)+2\epsilon_X 
\end{equation} where the maximal prior probability difference is
\begin{equation}
\epsilon_X= \underset{x,\tilde{x} \in \mathcal{D}^n : x \sim \tilde{x}}{\max} \ln \frac{p_X(x)}{p_X(\tilde{x})}
\end{equation} for neighboring datasets $x$ and $\tilde{x}$. This mutual information based mechanism satisfies $\epsilon-$ differential privacy.
In light of \cite[Theorem 1]{CY16}, which is visited in Section \ref{sec:shannon_inf}, the exact relation and ordering between conditional mutual information and differential privacy are today known.

\cite{PG21} studies the convergence of the source distribution estimate to the actual distribution based on the output from a locally differentially private mechanism. The fundamental difference in this setting stems from the fact that the differential privacy noise that is applied on each user's data locally, removes the requirement for a notion of neighborhood between datasets. In the model of \cite{PG21}, the source $\{X_i\}$ follows a discrete distribution $P$ and the mechanism $\mathcal{M}$ refers to the application of local differential privacy noise on $n$ i.i.d. source symbols which outputs the privatized observations $\{Y_i\}$ following the distribution $Q$, that is $P \mathcal{M}$. The goal of the legitimate observer is to estimate the source distribution $P$ using the noisy outputs $\{Y_i\}$ subject to either of f-divergence, mean-squared error (MSE) or total variation as the fidelity criteria. At the same time, an adversary aims to discover some source samples $X_i$. The authors present upper and lower bounds on their formulation of the trade-off between differential privacy level and fidelity loss based on the aforementioned three loss functions. 

\subsection{An Adaptation to Adversarial Classification}

Introducing adversarial examples to ML systems is a specific type of sophisticated and powerful attack, whereby additional (sometimes specially crafted) or modified inputs are provided to the system with the intent of being misclassified by the model as legitimate.
Adversarial classification is one possible defense proposed to correctly detect adversarial examples that aim to fool the classifier which detects outliers. In \cite{r17, UO22}, differential privacy is weaponized by the adversary in order to ensure to remain undetected. 
In addition to the statistical approach using hypothesis testing to establish a threshold of detection for the adversary as a function of the privacy budget, \cite{UO22} also introduces an original adaptation of lossy source coding to upper bound the impact of the attack. 

In this setting, the adversary not only wants to discover the data but also aims to harm the differentially private mechanism by modifying the released information without being detected. This trade-off between two conflicting goals of adversary is remodeled via the rate-distortion theory balancing the adversary's advantage and the security of the Gaussian differential privacy mechanism. 
Accordingly, the mutual information between the input and output of a communication channel in the original rate-distortion problem is now replaced by the datasets before and after the alteration applied by the adversary which are considered as neighbors, where the absolute difference between the two corresponds to the impact of the attack. Neighboring input vectors $X^n=\{X_1, \cdots, X_n\}$ and $\tilde{X}^n=\{X_1, \cdots, X_i, \cdots, X_n+X_{adv}\}$ are assumed to be i.i.d following the Gaussian distribution with the parameters $\mathcal{N}(0, \sigma^2_{X_i})$ with the difference of a single record denoted $X_{adv} \sim \mathcal{N}(0,\sigma^2_{adv})$. The query function takes the aggregation of this dataset as $q(\mathbf{X})=\sum_i^n X_i$ and the DP-mechanism adds Gaussian noise $Z$ on the query output leading to the noisy output in the following form $\mathcal{M}(\mathbf{X}, q(.), \epsilon, \delta)= Y = \sum_i^n X_i +Z$. An adversary adds a single record denoted $X_{adv}$ to this dataset. The modified output of the DP-mechanism becomes $\sum_i^n X_i + X_{adv} +Z$. 
\begin{theorem}
The privacy-distortion function for a dataset $X^n$ and Gaussian mechanism as defined by Definition (\ref{def:gauss}) is
\begin{equation}
P(s)=\frac{1}{2} \log\left( f_n \left( 1+\prod_i^n \sigma_{X_i}^{2}/s^2 \right)\right),
\end{equation} for $s\in \left[0, \prod_i^n \sigma_{X_i}^{2}\right]$ and zero elsewhere. $\sigma_{X_i}$ denotes the standard deviation of $X_i$ for $i=1,\cdots,n$, $f_n$ is some constant dependent on the size of the dataset $n$. %and $\sigma_{X_i}$. % is the standard deviation of the additional data.
\end{theorem} The sketch of the proof proceeds as follows.
The mutual information between the datasets before and the attack is derived as follows
\begin{align}
I(X^n;\tilde{X}^{n}) &= h(\tilde{X}^{n}) - h(\tilde{X}^{n}|X^n)\\
&\geq \frac{1}{2}  \sum_{i=1}^n\log \left(\left(2 \pi \mathrm{e}\right) \sigma^2_{X_i}\right) -\frac{1}{2} \log \left(2 \pi \mathrm{e} s^2\right) \\
&=\frac{1}{2} \log \left((2 \pi \mathrm{e})^{n-1}\prod_i^n \sigma^{2}_{X_i}/s^2\right) \label{eq:1st_expansion}
\end{align} %where $\sigma_{X_a}^{2}$ denotes the additional data $X_a$'s variance and the query function takes the aggregation of the entire dataset as $\sum_i^n X_i$. 
\begin{corollary} \label{cor:upper_bound}
The second order statistics of the additional data inserted into the dataset by the adversary is upper bounded as follows
\begin{equation}\label{upper_bound_attack}
\sigma^2_{X_{adv}} \leq \frac{1}{(2\pi \mathrm{e})^{n-1}}\left[\frac{s^2}{1-s^2/\sigma^2_{X_n}}\right] 
%\frac{\sum_i^n \sigma^{2}_{X_i} }{\left(\sum_i^n \sigma^{2}_{X_i}(2\pi \mathrm{e})^{n-1}/ s^2 \right)-1}
\end{equation} for $s^2= \frac{\sigma_z^2 \epsilon^2}{2 \log(1.25/ \delta)}$ and $n\geq 2$.
\end{corollary}
We have the following considering the neighbor that includes $X_{adv}$ has now $(n+1)$ entries over $n$ rows as $\tilde{X}^{n}=\{X_1, X_2, \cdots, X_n+X_{adv}\}$. Accordingly, the second expansion is derived on $X^n$ as 
\begin{align}
I(X^n;\tilde{X}^{n})&= h(X^n)-h(X^n|\tilde{X}^{n})\\
& \leq  \sum_{i=1}^{n}\frac{1}{2} \log \left(2 \pi \mathrm{e}\right)^n \sigma^2_{X_i}-\frac{1}{2} \log \left((2 \pi \mathrm{e})^n \sigma_{X_{adv}}^{2}\right)\label{eq11} \\
&\leq\frac{1}{2} \log \prod_{i=1}^{n-1} \sigma^2_{X_i}\left( 1+\frac{\sigma_{X_n}^{2}}{\sigma_{X_{adv}}^{2}}\right) \label{eq:2nd_expansion}
\end{align} leading to the upper bound in (\ref{upper_bound_attack}). Due to the adversary's attack, in the first term of (\ref{eq11}), we add up the variances of $(n+1)$ $X_i$'s including $X_{adv}$.
Since (\ref{eq:1st_expansion}) $\geq$ (\ref{eq:2nd_expansion}), we obtain the upper bound in Corollary \ref{cor:upper_bound}
For detailed derivation of (\ref{eq:1st_expansion}) and (\ref{eq:2nd_expansion}), the reader is referred to \cite{UO22}.
\begin{remark}The second expansion of the mutual information between neighboring datasets derived in (\ref{eq:2nd_expansion}), can be related to the well-known \textbf{rate-distortion function of the Gaussian source} which, originally, provides the minimum possible transmission rate for a given distortion balancing (mostly for the Gaussian case) the squared-error distortion with the source variance.
Combining (\ref{eq:1st_expansion}) with (\ref{eq:2nd_expansion}) characterizes the privacy-distortion trade-off of the Gaussian mechanism and bounds the impact of the adversary's modification on the original data in order to avoid detection in some sense \textit{calibrating} the adversary' attack to the sensitivity of the differentially private mechanism.\end{remark}

\section{ What \textit{else} do we want to learn?\label{sec:future}}

As convenient and practical it is, using information-theoretic quantities as privacy constraint is not fully exploited. This final part is reserved for concluding the tutorial by pointing out possible research directions on information-theoretic approaches to differential privacy for the future. 

In particular, for classification of adversarial examples in differentially private mechanisms where adversaries may seek for ways to harm the systems via modifying the ML model and misclassifying to model inputs, the source-coding
theory could provide new insights in the differential privacy measure itself. A great majority of the existing information theory literature benefits from
source coding theory for quantifying the privacy guarantee or for determining the leakage as already mentioned in Section \ref{sec:source_coding}. 
\cite{PG21} stands out in the way the rate-distortion perspective is translated for differential privacy where various fidelity criteria is set to determine how fast the empirical distribution converges to the actual source distribution. 
This approach could be extended for detection of adversarial examples attacking differentially private mechanisms beyond the work \cite{r17}, where the authors presented an application of the Kullback-Leibler differential privacy for detecting misclassification attacks in Laplace mechanisms. The corresponding distributions of relative entropy are considered as the differentially private noise with and without the adversary's advantage. The essential distinction that has to be made as relating differential privacy to mutual information is that the mutual information requires an input distribution. Differential privacy, on the other hand, is a characteristic of the mapping function applied on the input. Consequently, the query mechanism, hence the sensitivity, should play a role in defining the fidelity criterion as translating the adversarial classification into a rate-distortion problem similarly to \cite{UO22}.
Ultimately, this approach inspired by rate-distortion theory could be generalized beyond misclassification attacks for various types of attacks in order to determine and manipulate limits of the impact and detection probability of an attack, and to formally characterize a trade-off between the two. Moreover, by casting differential privacy for adversarial classification into a source coding problem, information-theoretic tools could be used to construct \textit{explicit coding strategies} for privacy preservation in anomaly detection.

Furthermore, information-theoretic quantities could shed light on connections between differential privacy and other trustworthy features of ML algorithms such as fairness and robustness. Various works show pairwise connections of differential privacy with robustness \cite{DJ20,PN20, PR19, r12} and fairness \cite{FT22}. The knowledge on these relations of differential privacy with these properties are yet to be explored from an information-theoretic perspective.

\bibliography{paper_adv_dp}

% Generated by IEEEtran.bst, version: 1.14 (2015/08/26)
\begin{thebibliography}{10}
\providecommand{\url}[1]{#1}
\csname url@samestyle\endcsname
\providecommand{\newblock}{\relax}
\providecommand{\bibinfo}[2]{#2}
\providecommand{\BIBentrySTDinterwordspacing}{\spaceskip=0pt\relax}
\providecommand{\BIBentryALTinterwordstretchfactor}{4}
\providecommand{\BIBentryALTinterwordspacing}{\spaceskip=\fontdimen2\font plus
\BIBentryALTinterwordstretchfactor\fontdimen3\font minus
  \fontdimen4\font\relax}
\providecommand{\BIBforeignlanguage}[2]{{%
\expandafter\ifx\csname l@#1\endcsname\relax
\typeout{** WARNING: IEEEtran.bst: No hyphenation pattern has been}%
\typeout{** loaded for the language `#1'. Using the pattern for}%
\typeout{** the default language instead.}%
\else
\language=\csname l@#1\endcsname
\fi
#2}}
\providecommand{\BIBdecl}{\relax}
\BIBdecl

\bibitem{S02}
L.~{S}weeney, ``k-anonymity: A model for protecting privacy,''
  \emph{International Journal of Uncertainty, Fuzziness and Knowledge-Based
  Systems}, vol.~10, pp. 557--570, 2002.

\bibitem{NS08}
A.~{N}arayanan and V.~{S}hmatikov, ``Robust de-anonymization of large sparse
  datasets,'' in \emph{IEEE Symposium on Security and Privacy}.\hskip 1em plus
  0.5em minus 0.4em\relax New York, NY, USA: IEEE, May 2008, pp. 111--125.

\bibitem{EJA+11}
K.~{E}mam, E.~{J}onker, L.~{A}rbuckle, and B.~{M}alin, ``A systematic review of
  re-identification attacks on health data,'' \emph{Plos One PMC}, vol.~6, pp.
  1--12, 2011.

\bibitem{ZB11}
H.~{Z}ang and J.~{B}olot, ``Anonymization of location data does not work: A
  large-scale measurement study,'' in \emph{Proc. Int. Conf. on Mobile
  Computing and Networking 17}.\hskip 1em plus 0.5em minus 0.4em\relax New
  York, NY, USA: ACM, Sep. 2011, pp. 145–--156.

\bibitem{C21}
\BIBentryALTinterwordspacing
A.~{C}hawla, ``Pegasus spyware -- a privacy killer,'' Jul 2021. [Online].
  Available: \url{SSRN}
\BIBentrySTDinterwordspacing

\bibitem{D06}
C.~{D}work, ``Differential privacy,'' in \emph{Automata, Languages and
  Programming}.\hskip 1em plus 0.5em minus 0.4em\relax Berlin, Heidelberg:
  Springer, 2006, pp. 1--12.

\bibitem{EPK14}
U.~{E}rlingsson, V.~{P}ihur, and A.~{K}orolova, ``{RAPPOR}: {R}andomized
  {A}ggregatable {P}rivacy-{P}reserving {O}rdinal {R}esponse,'' in \emph{ACM
  SIGSAC Conference on Computer and Communications}.\hskip 1em plus 0.5em minus
  0.4em\relax New York, NY, USA: ACM, Nov. 2014, pp. 1054--1067.

\bibitem{MKA+08}
A.~Machanavajjhala, D.~Kifer, J.~Abowd, J.~Gehrke, and L.~Vilhuber, ``Privacy:
  Theory meets practice on the map,'' in \emph{{IEEE} 24th International
  Conference on Data Engineering}.\hskip 1em plus 0.5em minus 0.4em\relax New
  York, NY, USA: IEEE, 2008, pp. 277--286.

\bibitem{D08}
C.~{D}work, ``Differential privacy: A survey of results,'' in
  \emph{International Conference on Theory and Applications of Models of
  Computation TAMC 2008, LNCS 4978}.\hskip 1em plus 0.5em minus 0.4em\relax
  Berlin, Heidelberg: Springer, 2008, pp. 1--19.

\bibitem{DS09}
C.~{D}work and A.~{S}mith, ``Differential privacy for statistics: What we know
  and what we want to learn,'' \emph{Journal of Privacy and Confidentiality},
  vol.~1, pp. 135--154, 2010.

\bibitem{DR05}
C.~{D}work and A.~{R}oth, ``The {A}lgorithmic {F}oundations of {D}ifferential
  {P}rivacy,'' \emph{Foundations and Trends in Theoretical Computer Science
  2014}, vol.~9, pp. 211--407, 2014.

\bibitem{LHA2}
Z.~Lin, M.~Hewett, and R.~Altman, ``Using binning to maintain confidentiality
  of medical data,'' in \emph{AMIA 2002 Annual Symposium Proceedings}, Nov.
  2002, pp. 454--458.

\bibitem{CY16}
P.~{C}uff and L.~{Y}u, ``{D}ifferential {P}rivacy as a {M}utual {I}nformation
  {C}onstraint,'' in \emph{{CCS} 2016, {V}ienna, {A}ustria}.\hskip 1em plus
  0.5em minus 0.4em\relax New York, NY, United States: Association for
  Computing Machinery, Oct. 2016, pp. 43--54.

\bibitem{WYZ16}
W.~{W}ang, L.~{Y}ing, and J.~{Z}hang, ``On the relation between
  identifiability, differential privacy and mutual information privacy,''
  \emph{IEEE Transactions on Information Theory}, vol.~62, pp. 5018--5029, Sep.
  2016.

\bibitem{M12}
D.~{M}ir, ``Information theoretic foundations of differential privacy,'' in
  \emph{International Symposium of Foundations on Practice of Security}.\hskip
  1em plus 0.5em minus 0.4em\relax Berlin, Heidelberg: Springer, Oct. 2012, pp.
  374--381.

\bibitem{BTV12}
M.~{B}ellare, S.~{T}essaro, and A.~{V}ardy, ``Semantic security for the wiretap
  channel,'' in \emph{Advances in Cryptology-CRYPTO}.\hskip 1em plus 0.5em
  minus 0.4em\relax Berlin, Heidelberg: Springer, 2012, pp. 294--311.

\bibitem{WXH21}
\BIBentryALTinterwordspacing
G.~{W}u, X.~{X}ia, and Y.~{H}e, ``Achieving dalenius' goal of data privacy with
  practical assumptions,'' May 2021. [Online]. Available:
  \url{https://arxiv.org/abs/1703.07474v5}
\BIBentrySTDinterwordspacing

\bibitem{r17}
A.~\"{U}nsal and M.~\"{O}nen, ``A {S}tatistical {T}hreshold for {A}dversarial
  {C}lassification in {L}aplace {M}echanisms,'' in \emph{IEEE Information
  Theory Workshop 2021}.\hskip 1em plus 0.5em minus 0.4em\relax New York, NY,
  USA: IEEE, Oct. 2021, pp. 1--6.

\bibitem{Pankratov-thesis}
D.~Pankratov, ``Communication complexity and information complexity,'' Ph.D.
  dissertation, The University of Chicago, Jun. 2015.

\bibitem{BJK+04}
Z.~{B}ar {Y}ossef, T.~{J}ayram, R.~{K}umar, and D.~{S}ivakumar, ``An
  information statistics approach to data stream and communication
  complexity,'' \emph{Journal of Computer and System Sciences}, vol.~68, pp.
  702--732, Jun. 2004.

\bibitem{MMP+10}
A.~{M}cGregor, I.~{M}ironov, T.~{P}itassi, O.~{R}eingold, K.~{T}alwar, and
  S.~{V}adhan, ``The limits of two-party differential privacy,'' in \emph{51st
  Annual Symposium on Foundations of Computer Science (FOCS)}.\hskip 1em plus
  0.5em minus 0.4em\relax 1730 Massachusetts Ave., NW Washington, DC, United
  States: IEEE Computer Society, Oct. 2010, pp. 81--90.

\bibitem{LKS19}
P.~Lin, C.~Kuhn, T.~Strufe, and E.~Jorswieck, ``Physical layer privacy in
  broadcast channels,'' in \emph{2019 IEEE International Workshop on
  Information Forensics and Security (WIFS)}.\hskip 1em plus 0.5em minus
  0.4em\relax Delft, Netherlands: IEEE, Dec. 2019, pp. 1--6.

\bibitem{M12+}
D.~{M}ir, ``Differentially-private learning and information theory,'' in
  \emph{International Workshop on Privacy and Anonymity in the Information
  Society PAIS}.\hskip 1em plus 0.5em minus 0.4em\relax New York, NY, USA: ACM,
  Mar. 2012, pp. 206--210.

\bibitem{HR22}
\BIBentryALTinterwordspacing
C.~{H}irche, C.~{R}ouz\'e, and D.~{F}ran\c{c}a, ``Quantum differential privacy:
  An information theory perspective,'' Feb. 2022. [Online]. Available:
  \url{https://arxiv.org/abs/2202.10717}
\BIBentrySTDinterwordspacing

\bibitem{AS19}
\BIBentryALTinterwordspacing
S.~Aaronson and G.~N. Rothblum, ``Gentle measurement of quantum states and
  differential privacy,'' in \emph{Proceedings of the 51st Annual ACM SIGACT
  Symposium on Theory of Computing}, ser. STOC 2019.\hskip 1em plus 0.5em minus
  0.4em\relax New York, NY, USA: Association for Computing Machinery, 2019, p.
  322–333. [Online]. Available: \url{https://doi.org/10.1145/3313276.3316378}
\BIBentrySTDinterwordspacing

\bibitem{ZM17}
L.~Zhou and M.~Ying, ``Differential privacy in quantum computation,''
  \emph{2017 IEEE 30th Computer Security Foundations Symposium (CSF)}, pp.
  249--262, 2017.

\bibitem{MM17}
M.~Senekane, M.~Mafu, and B.~Taele,
  ``\BIBforeignlanguage{English}{Privacy-preserving quantum machine learning
  using differential privacy},'' in \emph{\BIBforeignlanguage{English}{2017
  IEEE AFRICON}}, ser. 2017 IEEE AFRICON: Science, Technology and Innovation
  for Africa, AFRICON 2017, D.~Cornish, Ed.\hskip 1em plus 0.5em minus
  0.4em\relax United States: Institute of Electrical and Electronics Engineers
  Inc., Nov. 2017, pp. 1432--1435, iEEE AFRICON 2017 ; Conference date:
  18-09-2017 Through 20-09-2017.

\bibitem{DMNA06}
C.~{D}work, F.~{M}cSherry, K.~{N}issim, and A.~{S}mith, ``Calibrating {N}oise
  to {S}ensitivity in {P}rivate {D}ata {A}nalysis,'' in \emph{Theory of
  Cryptography Conference}.\hskip 1em plus 0.5em minus 0.4em\relax
  International Association for Cryptologic Research, 2006, pp. 265--284.

\bibitem{GV14}
Q.~{G}eng and P.~{V}iswanath, ``The optimal mechanism in differential
  privacy,'' in \emph{IEEE International Symposium on Information
  Theory}.\hskip 1em plus 0.5em minus 0.4em\relax New York, NY, USA: IEEE, Jul.
  2014, pp. 2371--2375.

\bibitem{DRV10}
C.~{D}work, G.~{R}othblum, and S.~{V}adhan, ``Boosting and differential
  privacy,'' in \emph{51st Annual Symposium on Foundations of Computer Science
  (FOCS)}.\hskip 1em plus 0.5em minus 0.4em\relax 1730 Massachusetts Ave., NW
  Washington, DC, United States: IEEE Computer Society, Oct. 2010, pp. 51--60.

\bibitem{KOV15}
P.~{K}airouz, S.~{O}h, and P.~{V}iswanath, ``The composition theorem for
  differential privacy,'' in \emph{32nd International Conference on Machine
  Learning}.\hskip 1em plus 0.5em minus 0.4em\relax JMLR, Inc. and Microtome
  Publishing (United States), 2015, pp. 4037--4049.

\bibitem{CF12}
F.~du~{P}in Calmon and N.~{F}awaz, ``Privacy against statistical inference,''
  in \emph{Fiftieth Annual Allerton Conference}.\hskip 1em plus 0.5em minus
  0.4em\relax New York, NY, USA: IEEE, Oct. 2012, pp. 1401--1408.

\bibitem{BK11}
G.~{B}arthe and B.~{K}\"opf, ``Information-theoretic bounds for differentially
  private mechanisms,'' in \emph{Computer Security Foundations
  Symposium}.\hskip 1em plus 0.5em minus 0.4em\relax New York, NY, USA: IEEE,
  2011, pp. 191--204.

\bibitem{AA11}
M.~S. Alvim, M.~E. Andr{\'e}s, K.~Chatzikokolakis, P.~Degano, and
  C.~Palamidessi, ``Differential privacy: On the trade-off between utility and
  information leakage,'' in \emph{Formal Aspects of Security and Trust}.\hskip
  1em plus 0.5em minus 0.4em\relax Berlin, Heidelberg: Springer Berlin
  Heidelberg, 2012, pp. 39--54.

\bibitem{D12}
A.~{D}e, ``Lower bounds in differential privacy,'' in \emph{Theory of
  Cryptography Conference}.\hskip 1em plus 0.5em minus 0.4em\relax
  International Association for Cryptologic Research, Mar. 2012, pp. 321--338.

\bibitem{BB+10}
B.~{B}arak, M.~{B}raverman, X.~{C}hen, and A.~{R}ao, ``How to compress
  interactive communication,'' in \emph{42nd ACM Symposium on Theory of
  Computing}.\hskip 1em plus 0.5em minus 0.4em\relax New York, NY, USA: ACM,
  Jun. 2010, pp. 67--76.

\bibitem{EGS3}
A.~{E}vfimievski, J.~{G}ehrke, and R.~{S}rikant, ``Limiting privacy breaches in
  privacy preserving data mining,'' in \emph{22nd ACM Symposium on Principles
  of Database Systems}.\hskip 1em plus 0.5em minus 0.4em\relax New York, NY,
  USA: ACM, Jun. 2003, pp. 211--222.

\bibitem{AC10}
M.~S. Alvim, K.~Chatzikokolakis, P.~Degano, and C.~Palamidessi, ``Differential
  privacy versus quantitative information flow,'' \emph{ArXiv}, vol.
  abs/1012.4250, 2010.

\bibitem{ZLV09}
S.~{Z}hou, K.~{L}igett, and L.~{W}asserman, ``Differential privacy with
  compression,'' in \emph{IEEE International Symposium on Information Theory,
  ISIT}.\hskip 1em plus 0.5em minus 0.4em\relax New York, NY, USA: IEEE, Jun.
  2009, pp. 2718--2722.

\bibitem{PG21}
A.~{P}astore and M.~{G}astpar, ``Locally differentially private randomized
  response for discrete distribution learning,'' \emph{Journal on Machine
  Learning Research}, vol.~22, pp. 1--56, Jul. 2021.

\bibitem{PKS20}
A.~{P}adakandla, P.~{K}umar, and W.~{S}zpankowski, ``Trade-off between privacy
  and fidelity vie ehrhart theory,'' \emph{IEEE Transactions on Information
  Theory}, vol.~66, pp. 2549--2569, Apr. 2020.

\bibitem{UO22}
\BIBentryALTinterwordspacing
A.~\"Unsal and M.~\"Onen, ``Calibrating the attack to sensitivity in
  differentially private mechanisms,'' \emph{Journal of Cybersecurity and
  Privacy}, vol.~2, no.~4, pp. 830--852, 2022. [Online]. Available:
  \url{https://www.mdpi.com/2624-800X/2/4/42}
\BIBentrySTDinterwordspacing

\bibitem{DJ20}
M.~Du, R.~Jia, and D.~Song, ``Robust anomaly detection and backdoor attack
  detection via differential privacy,'' in \emph{International Conference on
  Learning Representations ICLR 2020}, Sep. 2020.

\bibitem{PN20}
N.~Phan, M.~T. Thai, H.~Hu, R.~Jin, T.~Sun, and D.~Dou, ``Scalable differential
  privacy with certified robustness in adversarial learning,'' in
  \emph{Proceedings of the 37th International Conference on Machine Learning},
  ser. ICML'20.\hskip 1em plus 0.5em minus 0.4em\relax JMLR.org, 2020.

\bibitem{PR19}
\BIBentryALTinterwordspacing
R.~Pinot, F.~Yger, C.~Gouy-Pailler, and J.~Atif, ``{A unified view on
  differential privacy and robustness to adversarial examples},'' in
  \emph{{Workshop on Machine Learning for CyberSecurity at ECMLPKDD 2019}},
  Wurzburg, Germany, Sep. 2019. [Online]. Available:
  \url{https://hal.science/hal-02892170}
\BIBentrySTDinterwordspacing

\bibitem{r12}
M.~{L}ecuyer, V.~{A}tlidakis, R.~{G}eambasu, D.~{H}su, and S.~{J}ana,
  ``Certified robustness to adversarial examples with differential privacy,''
  in \emph{IEEE Symposium on Security and Privacy, San Francisco CA, USA}, May
  2019, pp. 1054--1067.

\bibitem{FT22}
\BIBentryALTinterwordspacing
F.~Fioretto, C.~Tran, P.~Van~Hentenryck, and K.~Zhu, ``Differential privacy and
  fairness in decisions and learning tasks: A survey,'' in \emph{Proceedings of
  the Thirty-First International Joint Conference on Artificial Intelligence,
  {IJCAI-22}}, L.~D. Raedt, Ed.\hskip 1em plus 0.5em minus 0.4em\relax
  International Joint Conferences on Artificial Intelligence Organization, 7
  2022, pp. 5470--5477, survey Track. [Online]. Available:
  \url{https://doi.org/10.24963/ijcai.2022/766}
\BIBentrySTDinterwordspacing

\end{thebibliography}
\bibliographystyle{IEEEtran}

\end{document}